# Rheology and microstructure of unsaturated granular materials: Experiments and simulations

M. Badetti, A. Fall, F. Chevoir, P. Aimedieu, S. Rodts and J.-N. Roux

Université Paris Est, Laboratoire Navier (UMR 8205 CNRS, IFSTTAR, Ecole des Ponts ParisTech),

2 Allée Kepler, Cité Descartes, F-77420 Champs-sur-Marne

Date : 26/02/18

**Abstract**

When dealing with unsaturated wet granular materials, a fundamental question is: what is the effect of capillary cohesion on the bulk flow and yield behavior? We investigate the dense-flow rheology of unsaturated granular materials through experiments and discrete element simulations of homogeneous, simple annular shear flows of frictional, cohesive, spherical particles. Dense shear flows of dry, cohesionless granular materials exhibit three regimes: quasistatic, inertial, and intermediate (Andreotti et al. 2013). Herewith, we show that the quasistatic and the intermediate regimes persist for unsaturated materials and that the rheology is essentially described by two dimensionless numbers: the reduced pressure $P^*$ comparing the cohesive to confining forces and the inertial number $I$, for a wide range of liquid content. This is consistent with recent numerical simulations (Khamseh et al. 2015). Finally, we measure the effective friction coefficient and the solid packing fraction variation throughout the wet bed. From this we show that, in the quasistatic regime, the Mohr-Coulomb yield criterion is a good approximation for large enough $P^*$. The experimental results agree quantitatively with the numerical simulations ones provided the intergranular friction coefficient $\mu$ is set to its physical value identified from dry material rheology (Badetti et al. 2018). To directly and quantitatively determine what happens inside the sheared granular bed, X-ray tomography measurements are carried out in a custom-made setup that enables imaging of a wet granular material after different shear histories. For the explored range of liquid content, samples remain homogeneous but exhibit some complex microscopic morphologies far from simple capillary bridges. From the X-ray microtomographic images, we can clearly distinguish liquid capillary bridges and liquid clusters by their morphologies. We see that that total number of capillary bridges decreases when one increases the liquid content and interestingly increases, at the expense of other morphologies, when we increases the shear strain. This latter explain the concordance between the experimental and numerical measurements since numerical model is restricted to the *pendular state*, for which the liquid phase is completely discontinuous and no coalescence occurs between liquid bridges.

I. INTRODUCTION

One of the most interesting properties of granular systems is that the grains only interact in the contact points, which form a random network inside the material. However, many powder processing methods such as granulation or coating require humid environments. The presence of liquid affects the properties of granular materials and the behaviour drastically depends on 'how wet' the grains are (Mitarai and Nori 2006). Indeed, it is commonly known that the addition of a small amount of liquid in a granular medium creates cohesion properties due to the surface tension of the liquid that wets the grains. Four basic states of wet granular material have been identified (Iveson et al. 2001; Mitarai and Nori 2006): *pendular*—liquid bridges between the contact points of the grains; *funicular*—both liquid bridges and liquid-filled pores; *capillary*—almost all the pores filled with liquid; and *slurry*—grains are fully immersed, no capillary action. Generally, these states can be distinguished by the level of liquid content. Such a mixture – unsaturated wet granular materials – may have a strong solidlike behavior (Herminghaus 2005), and, for instance, enable the building of sand castles as opposed to dry sand which cannot stabilize under gravity, with slopes steeper than the angle of repose (GDR Midi 2004; Pakpour et al. 2012).

In the dry case, the rheology is solely dictated by momentum transfer and energy dissipation occurring in direct contacts between grains and with the walls (Andreotti et al. 2013). Despite the simplicity of the system, the behaviour of dry granular materials is very rich and a major step toward describing the rheological properties was the generalization of the Coulomb friction approach (Da Cruz et al. 2005; Jop et al. 2006). This relates the shear stress $\tau$ to the confinement pressure $\sigma_n$ via a friction coefficient $\mu_d^*$ that depends on the dimensionless inertial number $I$ (the $\mu(I)$–rheology) (GDR Midi 2004; Da Cruz et al. 2005; Jop et al. 2006; Hatano 2007; Staron et al. 2010; Gray and Edwards 2014; Fall et al. 2015b).

The presence of a small amount of interstitial fluid in the system introduces another degree of complexity due to the cohesive forces between particles in addition to the friction force in dry granular matter. The liquid, in which the pressure is lower than in the void phase, creates adhesive forces. Their macroscopic effects, in the quasistatic regime, are traditionally described in terms of a cohesion, $c$, appearing in the Mohr-Coulomb condition as:

$$\tau = \mu_d^* \sigma_n + c \qquad (Eq.1)$$

in which $\mu_d^*$ the macroscopic friction coefficient of the dry grain assemblies (Pierrat et al. 1998). The Mohr-Coulomb stability criterion is analogous to the well-known Coulomb friction law for two sliding surfaces.

The strong influence of the capillary forces on the shear strength is perhaps the most reported property of wet granular materials. However, investigations are mostly limited to the quasistatic behaviours and the influence of shear rate on the mechanical properties is less studied. Bocquet et al. (Bocquet et al. 1998, 2002) studied the effect of waiting time on the repose angle of glass beads contained in a rotating tumbler, and found logarithmic aging of the maximum static angle.

Wet granular behaviours in different experimental devices have been studied, for example, in a rotating drum (Tegzes et al. 2003), in a shear apparatus (Chou and Hsiau 2010), on an inclined plane (Samadani and Kudrolli 2000, 2001) and in a vibrating bed (Scheel et al. 2004). The system starts to flow when the externally imposed stress exceeds the inter-aggregate contact forces (Day and Holmgren 1952). It was found that a transition from a gaseous regime to a viscoplastic regime occurs when the liquid content reaches a critical point, which is dependent on the particle size (Geromichalos et al. 2003). In wet granular systems, liquid bridges are formed by small regions of liquid in the contact area of particles, in which due to surface tension effects a low capillary pressure prevails. Existing studies thus indicate that cohesion strongly affects the behaviour of dense granular flow as well as its microstructure: the mechanical properties at small liquid content are determined by the liquid bridges between grains, and those at larger liquid content are determined by the flow of the liquid through the pores (Ghezzehei and Or 2000). Besides, the characterization of flow properties is very different; these materials are reported to exhibit a mixed behavior of elasticity, viscosity, and plasticity (Tegzes et al. 2003; Mitarai 2012). The reason is that, the few existing studies were done in setups in which shear banding (localization) and avalanches strongly influence the apparent rheology, and so it is not obvious that results for a given setup can be translated to a generally applicable conclusion.

In order to attain a general understanding of the role of cohesion in granular materials, it was shown from Discrete Element Method simulations (DEM) that the internal state of the material, in shear flow under controlled normal pressure $\sigma_n$ (Rognon et al. 2006, 2008; Berger et al. 2015; Khamseh et al. 2015; Than et al. 2017), depends on two dimensionless parameters: a reduced pressure $P^*$ (similar to the "cohesion number" which was defined as $\eta = 1/P^*$ in (Rognon et al. 2008; Berger et al. 2015)) comparing the cohesive to confining forces and an inertial number defined as $I = \dot{\gamma} d_p / \sqrt{\sigma_n / \rho_p}$.

The simplest model of a granular material in the presence of a wetting liquid at small amount assumes the liquid to be confined in menisci at contacts or in narrow gaps between neighbouring grains (Scheel et al. 2008). The pairwise attractive force that stems from such liquid bridges is well known (Iveson et al. 2001). It was implemented in DEM simulations (Mikami et al. 1998; Gröger et al. 2003; Richefeu et al. 2006; Scholtès et al. 2009; Khamseh et al. 2015), resulting in good agreements with macroscopic experimental results (Chateau et al. 2002; Richefeu et al. 2006; Soulié et al. 2006) of some quasistatic properties. For monodisperse grains, the adhesion force $F_0$ is the only internal force and it should be compared to the repulsive contact forces induced by the confining stress. Hence, the cohesion of the system is characterized by $P^*$ defined as:

$$P^* = \frac{\sigma_n d_p^2}{F_0} \approx \frac{\sigma_n d_p}{\pi \Gamma} \qquad (Eq.2)$$

where $F_0 = \pi d_p \Gamma$ is the maximum capillary force between a pair of grains (assuming that the contact angle of the liquid with the surface of the beads is small) and $\Gamma$ the liquid surface tension. For our experimental system: surface tension of silicon oil, $\Gamma$ = 20.6 mN/m, particle diameter $d_p$ = 0.5 mm and at a normal stress $\sigma_n$ =129.4 Pa, one gets $P^*$ = 1. The case $P^* \to +\infty$ corresponds to cohesionless systems, and when $P^* \gg 1$, the confining forces are dominant and the effects of cohesive forces become negligible (Khamseh et al. 2015). Thus $P^*$ is a critical microscopic parameter that controls the macroscopic rheology of the system. Thereby $P^*$ extends the rheology of inertial flows to cohesive granular materials (Khamseh et al. 2015).

From DEM simulations, it has been thus observed (Gilabert et al. 2008; Than et al. 2017) that with different cohesive granular models, cohesive forces are expected to have strong effects, possibly enabling very loose equilibrium microstructures for $P^* \ll 1$, while the properties of cohesionless systems are retrieved in the limit of large $P^*$. Note that results in 2D as well as in 3D simulations, lower $P^*$ values increase the internal friction coefficient $\mu_w^* = \tau/\sigma_n$ and decrease the solid fraction $\phi_S$ of the wet sheared granular material (unsaturated granular material). However, the effect on $\mu_w^*$ is considerably larger in the assembly of wet particles in three dimensions: even for $P^* = 1$ (Rognon et al. 2006, 2008; Berger et al. 2015; Khamseh et al. 2015). Despite the fact that those simulations were restricted to the *pendular* regime (independent menisci) and used simple rules for liquid distribution (e.g., constant meniscus volume), some straightforward extensions are possible, as regards, in particular, the rules governing liquid transport and exchange between different menisci (Mani et al.

2013). In the same time, viscous effects, which affects the interaction law and might strongly affect the dynamics (Lefebvre and Jop 2013), have not been included into simulations. Nevertheless, they revealed a wealth of remarkable phenomena extending the rheology of dry granular to cohesive granular materials.

At higher liquid content, inter-granular liquid bridges merge, as observed in X-ray microtomography (Herminghaus 2005; Fall et al. 2014; Saingier et al. 2017), thereby undermining the adequacy of the binary interaction model as implemented, e. g., in (Richefeu et al. 2006), with its simple rules for the spatial distribution of the liquid. It has been found a rich variety of liquid cluster morphologies (Scheel et al. 2008) beyond the well-studied liquid bridge regime. The number and the size of observed liquid clusters strongly depend on the liquid content. Herminghaus (Herminghaus 2005) showed that the average number of liquid bridges increased with the increasing liquid content, and reached a stable value when the liquid content exceeded a critical value. The self-diffusion coefficients and the fluctuation velocities also decrease as the liquid content and liquid viscosity in the wet granular system increase (Chou and Hsiau 2010). As liquid content increases further, the pores among grains are progressively filled and at some point the system becomes a dense suspension, where both grain-liquid and grain-grain interactions play important roles (Iveson et al. 2001). The basic physics is that shear strength starts to increase when capillary bridges form between the grains. For increasing liquid fraction, the capillary bridges merge and eventually disappear when the granular media is fully saturated. There must be a maximum strength at a finite amount of added liquid. Indeed, the magnitude of the cohesive forces at different liquid content is intimately linked to the morphology of the interstitial fluids on the scale of single grains (Fall et al. 2014; Semprebon et al. 2016). Bruchon and co-workers used X-ray Computer Tomography (CT) to analyze changes in the three-dimensional grain skeleton structure upon wetting a sand (Bruchon et al. 2013). In their tests, a small cylindrical specimen of loose sand was wetted from the bottom and simultaneously scanned in a CT-device. The results indicated that the local collapse behaviour is related to the coalescence of capillary bridges in the grain skeleton upon wetting. However, investigations of the influence of cohesion on the flow property of wet granular materials are mostly limited to the quasistatic regime and the influence of shear rate on the mechanical properties in relation to the microstructure does not exist to our knowledge.

In this paper, we try to bring new insights on these issues by carefully analyzing the flow of unsaturated wet grains in the dense regime, following the work of (Rognon et al. 2008; Berger et al. 2015; Khamseh et al. 2015) but with an emphasis on the quantitative determination of the microstructure and rheological laws. In order to establish the constitutive laws of the flow of wet granular material, we impose a large enough strain rate to the sample to approach steady flow regime wherein we can measure macroscopic quantities such as shear stress and solid volume fraction for a given confining pressure. Experimental results are then compared to DEM simulations carried out on model wet granular material using the same method as in (Khamseh et al. 2015; Than et al. 2017). Moreover, to obtain the details of the grain packing geometry and the liquid distribution within the sheared sample, we also used a rheometer coupled to an X-ray microtomograph to characterize the sample' microstructure.

## II. MATERIALS AND METHODS

### 1. WET GRANULAR MATERIALS

The experiments are carried out on model materials: slightly polydisperse assemblies of macroscopic solid spherical beads, mixed with a non-volatile, wetting, Newtonian liquid. We use a granular material made of rigid polystyrene beads (from *Dynoseeds*) of density $\rho_p$ = 1050 kg/m³ and of diameter $d_p$ = 0.5 mm (with a standard deviation of 5%, sufficient to prevent crystallization). Subsequently, silicone oil (viscosity 50 mPa.s; density 0.95 g.cm$^{-3}$; surface tension $\Gamma$ = 20.6 mN/m and wetting angle 2° $\leq \theta \leq$ 5°) is mixed thoroughly with the dry beads, after which the system is poured into the annular shear cell (Fall et al. 2015b) and compacted by repeated tapping. We define the liquid content as $\varepsilon = V_\ell / V_s$, where $V_\ell$ and $V_s$ are the volumes of liquid and beads respectively. Different liquid contents were tested ($\varepsilon \in$ [0.015 – 0.075]) corresponding to the *pendular* state (Mitarai and Nori 2006). We use silicon oil rather than water since the polystyrene beads are not wettable by water. Note that we have checked by NMR (data not shown) that drainage occurs for $\varepsilon$ >0.075: the liquid flows towards the bottom of the sample. All experiments discussed below are carried out, at ambient conditions, by preparing a homogeneous material.

### 2. X-RAY MICROTOMOGRAPHY

To obtain the details of the grain packing and the liquid morphology within the granular pile, we carried out X-ray microtomography (X-ray CT) experiments. X-ray CT is an imaging tool frequently used in materials science to provide three-dimensional structural information of complex materials in a non-destructive way (Scheel et al. 2008; Bruchon et al. 2013). The measurements are conducted with an *Ultratom* scanner from *RX Solutions*, equipped with a *L10801 Hamamatsu* X-ray source (source parameters: 110 kV–125 µA) associated with a *Varian PaxScan* 2520V flat panel detector used at full resolution. 3D images encoding for the X-ray absorption field are reconstructed from the recorded 2D radiographs. Consistently with the definition of the final 3D images, 1440 projections were taken over 360° rotation with small rotation step. Note that for each rotation angle, 8 radiographs (with an exposure time for one radiograph of 0.25 s) have been averaged to improve the signal to noise ratio. The final 3D images had a spatial resolution (the voxel size of the image) of 14 μm and a definition of 1648x1294x1678 voxels. Note that, with our X-ray CT device and the chosen spatial resolution, the whole sample can be scanned in 45 minutes.

3. STEADY SHEAR EXPERIMENTS UNDER CONFINEMENT PRESSURE

Steady shear experiments were done using two types of *stress-controlled* rheometers (*Kinexus Pro* by *Malvern* and *MCR 502* by *Anton Paar*). The wet sample is poured and compacted by repeated tapping into an annular shear cell geometry (Fall et al. 2015b).

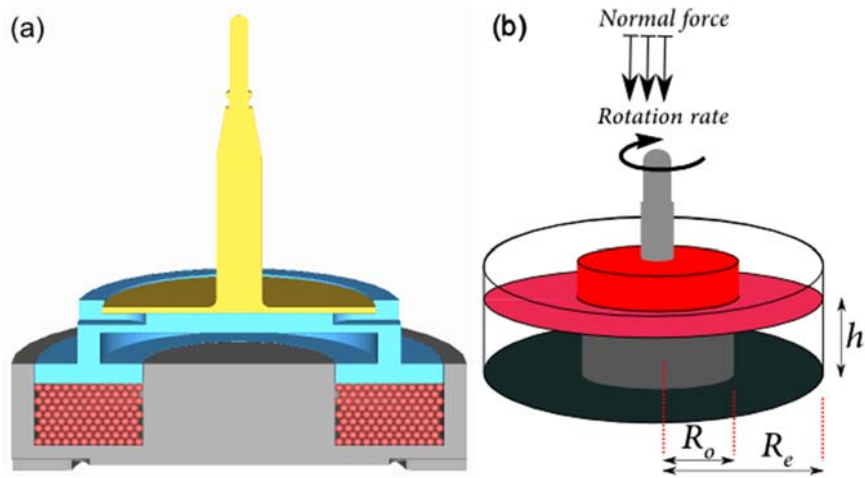

Figure 1: (a) and (b) Cross section of the annular plane shear flow. The shear and pressure are provided by a ring which is assembled on the rheometer that is free to move vertically while maintaining a constant rotation rate or shear rate and imposed pressure.

The normal-stress controlled annular apparatus was designed to carry out such experiments, in which the granular phase may dilate or contract, depending on the density necessary to support $\sigma_n$ under the imposed strain rate $\dot{\gamma}$. Instead of setting the value of the gap size for a given experiment, as in previous studies (S. B. Savage and M. Sayed 1984) and generally in rheometric measurements, we impose the normal force (i.e., the confining normal stress $\sigma_n$) and then, under shear, we let the gap size vary in order to maintain the desired value of the normal force. We then have access to instantaneous measurements of the driving torque $T$ and the gap $h$ variation for imposed normal force $F_n$ and $\dot{\gamma}$: in this case, the solid volume fraction $\phi_S$ is not fixed but adjusts to the imposed shear (Boyer et al. 2011; Kuwano et al. 2013; Fall et al. 2015b).

## 4. NUMERICAL SIMULATIONS

DEM simulations were carried out on model wet granular materials. As in (Khamseh et al. 2015; Than et al. 2017; Badetti et al. 2018), the simulated systems are assemblies of monodisperse spherical beads of diameter $d_p$, interacting in their contacts by Hertz-Mindlin elastic-frictional forces, with the elastic properties of polystyrene. Let us stress however, that the intergranular friction coefficient is fixed to $\mu = 0.09$ in order to capture the experimental data of polystyrene dry grains (Fall et al. 2015b; Badetti et al. 2018). Indeed in the quasistatic regime, numerical simulations show that both $\mu_d^*$ and $\phi_S$ of dry grains are functions of $\mu$: while the solid fraction is a decreasing function, the internal friction coefficient increases with increasing $\mu$ as previously reported (Lemaître et al. 2009). For wet granular materials, the meniscus volume is fixed in the pendular regime in which the liquid forms disjoint menisci bridging pairs of grains in contact or close to one another. For small volumes, liquid bridges introduced attractive capillary forces for which the Maugis model (Maugis 1987) is adopted for simulations. Assuming that the value of $\mu$ does not change in the presence of the wetting liquid (Boyer et al. 2011), the results are shown below depending of $\mu_w^*$ (the effective friction coefficient) and $\phi_S$ as a functions of $I$ and $P^*$ and compared to experimental data.

## III. RESULTS
### 1. IMAGE ANALYSIS AND LIQUID MORPHOLOGIES

The wet sample is poured and compacted by repeated tapping into a parallel plate in cup geometry (Fiscina et al. 2012) (of $h = 9$ mm of gap and $R_p = 10$ mm of radius) inserted in the X-ray microtomograph (Ovarlez et al. 2015). Subsequently, the three-dimensional gray scale images generated by the reconstruction algorithm were processed with *XAct* from *RX Solutions*. Due to the different X-ray absorption contrast of the three phases: liquid, beads and air, each one can be clearly distinguished from each other.

Horizontal slices through 3D tomographic images are shown in Fig. 2 for liquid contents of $\varepsilon = 0.015$, $\varepsilon = 0.03$, and $\varepsilon = 0.075$ respectively. At $\varepsilon = 0.015$ only capillary bridges can be found in the sample exclusively at the contacts. Whereas for a larger liquid content $\varepsilon = 0.03$, liquid clusters are visible. For an even larger liquid content $\varepsilon = 0.075$, larger clusters between many beads have formed.

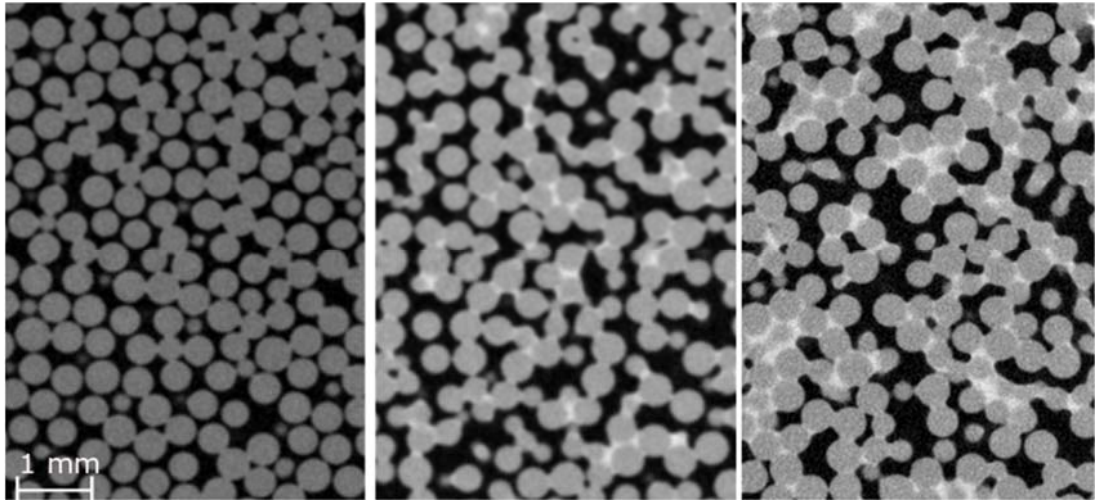

Figure 2: Horizontal slices through 3D tomographic images of randomly distributed polystyrene beads 500 μm diameter at various liquid contents $\varepsilon = 0.015$, $\varepsilon = 0.03$, and $\varepsilon = 0.075$ respectively. Liquid and air appears as white, respectively, black areas while the glass beads appear as gray discs. The slice cuts the beads at different heights and thus the beads appear to be more polydisperse than they actually are.

For a quantitative analysis of the tomography images, the raw data had to be smoothed (to reduce the noise of the images) and segmented (to constitute entities that correspond for example to the solid, gaseous and liquid phases in the case of an unsaturated granular). At each step, a wide range of existing tools can be used (Coster and Chermant 2001). In the first pre-processing step, we reduce the noise in each image by applying a bilateral filter and then "rolling ball" algorithm is used to correct for uneven illuminated background by simple thresholding method. The second step is the segmentation of images in which the grain and liquid phase were used respectively to extract the volume, the surface and the position of grains on one hand and the liquid content one the other hand. All further analyses of the segmented images were performed with custom-made programs using *Python programming language*. In this way, we obtained the statistics of the number of capillary bridges and liquid clusters in contact with a bead, as well as the volume of individual liquid clusters.

In the X-ray microtomographic images, we can clearly distinguish liquid capillary bridges and liquid clusters by their morphologies. The capillary bridge is the liquid morphology between two particles and a 3D capillary bridge is shown in Figure 3(a-b). The limited resolution of the X-ray microtomography results in deviations from the real shape of the capillary bridges. For example capillary bridges with two particles in contact can be resolved as 'donuts' with a hole in the center and the diameter of this hole will decrease as the voxel resolution is increased. It is clearly observed that the total number of capillary bridges decreases when one increases the liquid content. Above a critical liquid content 0.015 $< \varepsilon <$ 0.03, intergranular liquid bridges merge in clusters. The percolating cluster grows as the liquid content is further increased (Scheel et al. 2008). The smallest liquid clusters are here referred to 'dimer' and 'trimer'. They are formed between three beads: the dimer corresponds to two capillary bridges with one contact point between four particles as shown in Fig. 3(c), while the trimer has three contact points (Fig. 3(d)). Some larger clusters are illustrated in Fig. 3(e) – pentamer – cluster that has five contacts between four beads and in Fig. 3(f) a more larger one.

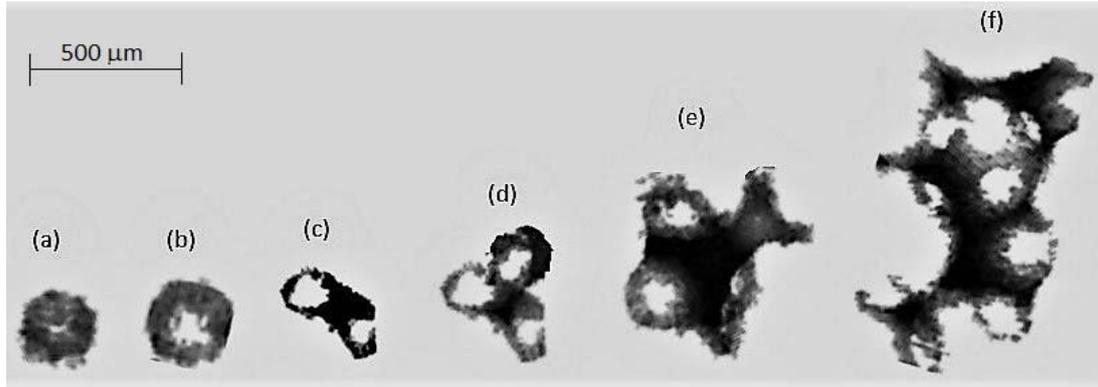

Figure 3: Shapes of some liquid morphologies: from capillary bridge to large clusters

To explore, how the number of capillary bridges and other morphologies on a bead varies with the shear strain, we prepared samples with a fixed liquid content: $\varepsilon =$ 0.03 and $\varepsilon =$ 0.075. The shear strain $\gamma$ was varied between 0 and 35. Note that, since shear is inhomogeneous in a parallel plate geometry, we restrict the analysis of liquid morphology to the region of the gap of radial position $0.45 R_p \leq r \leq 0.9 R_p$. In this region, shear is roughly homogeneous, and the microstructure can be studied as a function of the average shear strain $\gamma = \phi R_{mean} / h$ by changing the angular displacement $\phi$ of the upper plate of the geometry. Note that $R_{mean}$ is defined as $R_{mean} = (0.45 R_p + 0.9 R_p)/2$. Since particles and also the wetting fluid must not move during the time of a scan, this leads us to apply to our sample a given shear strain, then stop the flow and image the material at rest with a frozen structure, which is supposed to be close to the structure under flow (Ovarlez et al. 2015). The underlying hypothesis is that relaxation of the microstructure at rest is negligible. However, when the shear is stopped, a certain time is needed for the liquid morphologies to equilibrate (Scheel et al. 2008). During this process, small capillary bridges will grow on the expense of larger cluster. A time series of tomographies was recorded 10 min after the shear has been stopped. Figure 4 shows a histogram of the frequency of occurrence of liquid capillary bridges and other morphologies found in a granular pile with two liquid contents $\varepsilon = 0.03$ and $\varepsilon = 0.075$ for different values of shear deformation. A statistical analysis of the tomography data reveals that the number of capillary liquid bridges decreases as $\varepsilon$ increases further since capillary bridges merge into liquid clusters. Moreover, we observed that their frequency increases, at the expense of other morphologies, when we increases the shear strain.

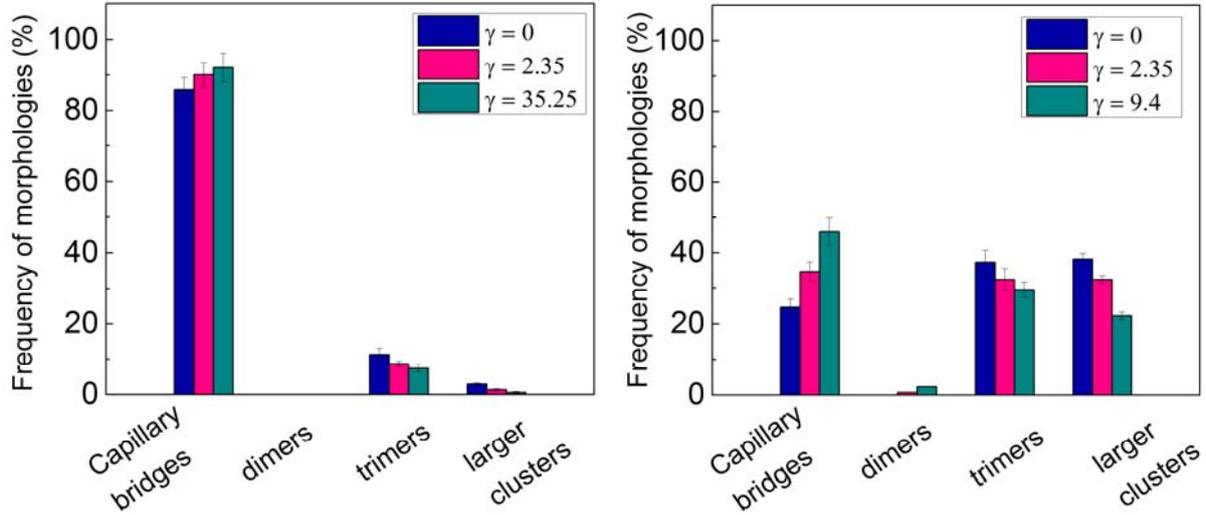

Figure 4: Histogram of the frequency of liquid morphologies as a function of shear deformation for two liquid contents $\varepsilon = 0.03$, and $\varepsilon = 0.075$ respectively.

2. CONSTITUTIVE LAWS: EXPERIMENTS VERSUS DEM SIMULATIONS

Typical experimental measurements are shown in Fig. 5, where we start out with a gap $h_0$, which represents typically few particle diameters ($10 d_p$ to $45 d_p$), impose a constant $\dot{\gamma}$ and $F_n$ and measure the torque $T$ and the gap $h$ as functions of strain (or time).

The system reaches a steady state after a certain amount of strain: the critical state is defined by the local shear accumulated over time under a constant confining pressure and constant strain rate condition. This state is reached after large enough strain, when the material deforms with applied shear without any change in state variables, independent of the initial condition. Note that dense materials show a systematic dilation while loose samples present a compaction before reaching their critical state (Fall et al. 2015b).

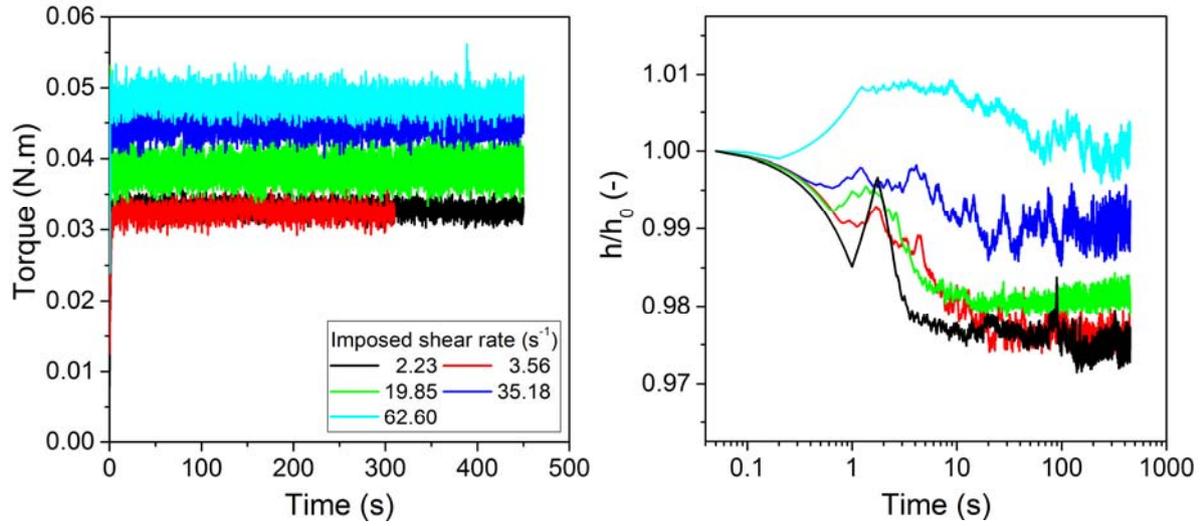

Figure 5: Evolution as a function of time at 2.96N imposed normal force under different applied shear rates of: (a) the driving torque and (b) the rescaled gap size. (Only few curves are shown for clarity)

Here we present general definitions of the averaged macroscopic quantities – including strain rate, stresses and the solid volume fraction. In the annular plate-cup shear geometry, the driving torque is related to the local value $\tau$ of the shear stress by:

$$T = 2\pi \int_{R_i}^{R_o} \tau\, r^2 dr \qquad (Eq.3)$$

where $R_i$ = 21 mm and $R_o$ = 45 mm are inner and outer radii of the annular trough. If the radial velocity gradient is neglected (Cleaver et al. 2000; Coste 2004) into the annular trough, the shear stress is quasi-independent of the radial position *r* and thus, integrating Eq. (3) yields the shear stress as:

$$\tau = 3T / 2\pi (R_o^3 - R_i^3) \qquad (Eq.4)$$

Equation (4) holds because the lateral contribution of wall frictions on the stress distribution within the granular sample can be neglected due to the lubrication from the wetting liquid (Boyer et al. 2011). Moreover, the cylinders of our annular shear cell were finished as smoothly as possible to permit the granular material to slip there as readily as possible (Fall et al. 2015b).

The normal stress can also be calculated from the normal force as follows:

$$\sigma_n = (F_n + mg/2) / \pi (R_o^2 - R_i^2) \qquad (Eq.5)$$

Note that for $h_0 \approx (10 - 45) d_p$, the imposed normal stress is larger than the hydrostatic pressure once $F_n$ is larger than 1.20 N, meaning that gravity may be neglected for the explored range of imposed normal forces.

Besides, assuming that the velocity gradient is approximately uniform over the depth and width of the annular trough and a no-slip condition exists at the rough upper and lower shearing walls, one can estimate the mean shear rate averaged across the annulus as:

$$\dot{\gamma} = \Omega(R_o + R_i)/2h \quad \text{(Eq.6)}$$

And the mean shear strain is given by:

$$\gamma = \varphi(R_o + R_i)/2h \quad \text{(Eq.7)}$$

where $\varphi$ and $\Omega$ are respectively the angular displacement and the rotation speed of the upper plate.

The gaps between the moving upper plate and the side vertical wall are smaller than 100 μm, which is five times as small as the grain size. Since these gaps are so narrow that grains cannot escape from the shear cell, one can measure unambiguously the solid fraction from the gap variation and the mass $m$ of grains as:

$$\phi_S = m/\pi\rho_p h(R_o^2 - R_i^2) \quad \text{(Eq.8)}$$

The vertical position $h(t)$ of the plate indicates the dilation or compaction of the granular material.

Defining the macroscopic friction coefficient of the unsaturated granular material as the ratio of the shear stress $\tau$ to the confining pressure $\sigma_n$ in steady state as $\mu_w^* = \tau/\sigma_n$, the macroscopic rheological response of the system of wet particles can be therefore presented in the form of non-dimensional macroscopic friction coefficient $\mu_w^*$ and solid fraction variations $\phi_S$ versus the non-dimensional inertial number $I$ for various $P^*$.

We report here the variation of $\mu_w^*$ and $\phi_S$, from both experiments and simulations, with $I$ for various $P^*$. For the two quantities, a good agreement is observed, once intergranular friction coefficient is set to its appropriate value. Figure 6(a) shows, for different values of $P^*$, how $\mu_w^*$ varies throughout the flow regimes. In the quasistatic limit ($I \leq 0.001$) which is strongly influenced by capillary forces: $\mu_{w,0}^* \equiv \mu_w^*(I \to 0)$ increases with decreasing $P^*$. When the inertial number is increased, $\mu_w^*$ increases whatever the reduced pressure. Moreover, all data points are above the curve of the cohesionless dry beads (in the absence of the capillary forces: $P^* \to \infty$).

Complementary information is the dynamical *dilatancy law*, which describes the variations of the solid fraction as a function of the inertial number: $\phi_S$ decreases with increasing $I$ (Fig. 6(b)). As for the friction coefficient, the solid fraction variation is strongly influenced by $P^*$: $\phi_S$ is an increasing function of $P^*$ for the explored range of inertia number. In the quasistatic regime, $\phi_{S;0} \equiv \phi_S(I \to 0)$ decreases from 0.615 to 0.605 when $P^*$ decreases.

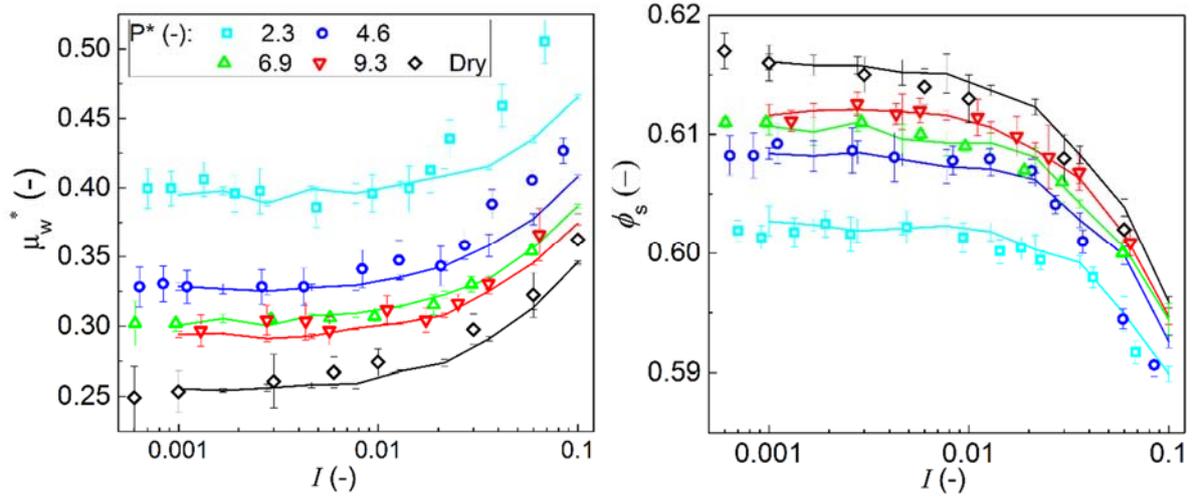

Figure 6: Experimental (symbols) and simulation (lines) results for macroscopic friction coefficient $\mu_w^*$ (a) and solid fraction $\phi_S$ (b) versus inertial number $I$ for different values of reduced pressure $P^*$. The error bars correspond to the standard deviation of the torque and the gap level in the steady state.

The increase of $\mu_w^*$ and the decrease of $\phi_S$ for moderate values of $I$ show the usual behavior of granular materials under shear flow, similar to other experimental and numerical studies on dry grains (GDR Midi 2004; Jop et al. 2006; Hatano 2007; Berger et al. 2015; Fall et al. 2015b; Khamseh et al. 2015). Note that however, for high values of $I$, the shear resistance is underestimated by the simulations: $\mu_w^*$ grows faster with $I$ in the experiments than in simulations. However, the critical inertial number (data not shown) marking the transition between the quasistatic and intermediate regimes decreases strongly as $P^*$ increases so that the friction coefficient (respectively, the solid fraction) increases (decreases) faster for the dry sample than for the wet ones. Indeed, as contacts are stabilized by attractive (capillary) forces that hold the granular system together, they do not so easily open when the network is being sheared that should limit the dilating tendency of faster flows.

Nevertheless, so far, only a few attempts have been made to answer the question about the effect of cohesion on the presence of shear banding: (Fall et al. 2010) concerning dense adhesive emulsions, (Estrada et al. 2010) on cemented granular media and (Schwarze et al. 2013; Khamseh et al. 2015) on wet granular media. In wet granular materials, Khamseh and co-workers have shown a permanent shear banding for low $P^* = 0.1$ and that the localization tendency of the flow increases for smaller values of inertial number. Rheological studies on adhesive and nonadhesive emulsions (Fall et al. 2010; Paredes et al. 2011) reported that the presence of attractive forces at contact affects shear banding by affecting flow heterogeneity and wall slip.

For wet granular materials, to establish constitutive laws of the flow, it is essential to ensure that the flow, at least from a macroscopic point of view, is homogeneous. To investigate that, we have varied the initial size of the gap from $10 d_p$ to $44 d_p$.

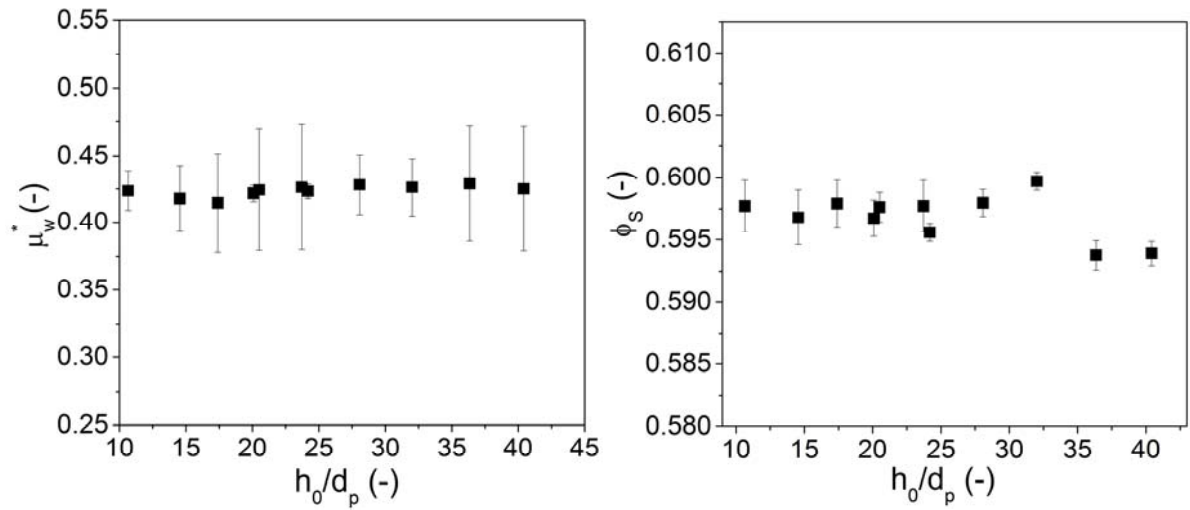

Figure 7: Macroscopic friction coefficient and solid fraction versus gap at $I$ =0.03 and $P^*$ = 2.3.

Figure 7 illustrates the evolution of the friction coefficient and of the solid fraction with the gap heights for $P^*$ = 2.3 and $I = 0.03$: it shows that changing the gap does not significantly affect these results. This suggests a total absence of shear localization at this reduced pressure. However, at small $P^* \leq 2$, the resolution of our measurements is not sufficient to dismiss the possibility that shear localization arises.

3. MACROSCOPIC COHESION

Previously reported experiments of unsaturated granular materials show that the tensile strength and the yield stress have a typical dependence on the wetting liquid fraction. The friction coefficient $\mu_{w,0}^*$ in the quasistatic regime increases with decreasing $P^*$ and increasing the liquid content (Badetti et al. 2018). If there is no geometric ordering of the system, it has been shown that these capillary forces introduce a predominantly isotropic compressive stress into the system (Pierrat et al. 1998). The effective stress of the wet granular material will then be in each case represented by a Mohr circle shifted to the right with both principal stresses increased by the value of the isotropic compressive stress. In the case of non-cohesive dry materials, the steady shear stress is a linear function of pressure, as predicted by the Mohr-Coulomb criterion with a slope that increases with increasing shear rate (Fall et al. 2015b). However, for wet granular materials, the relation between shear stress and confining pressure becomes non-linear when cohesion $c$ is introduced at the contacts due to capillary forces. Taking $c$ independent of $P^*$, (Eq. 1) implies that the friction coefficient $\mu_{w,0}^*$ should varies linearly with $1/P^*$ as:

$$\mu_{w,0}^* = \mu_{MC}^* + \frac{C^*}{P^*} \qquad (Eq.9)$$

where $C^* = cd_p / \pi\Gamma$ is a dimensionless cohesion and $\mu_{MC}^*$ the friction coefficient of the dry sample.

To evidence that, measurements were done in the quasistatic regime with samples subjected to a very low shear rate under different confining pressures corresponding to an inertia number of $I$ = 0.001. The resulting steady stresses and solid fraction were measured in the same way as discussed above. Figure 8 uses Eq. 9 to identify values of cohesion $C^*$. Indeed, the linearity of the graphs shows that the Mohr-Coulomb criterion fits well the experimental points, at least as long as $P^* \geq 2$. For smaller $P^*$, the shear resistance is overestimated by the Mohr-Coulomb criterion. Moreover, we evidence once again that when $P^* \to \infty$, wet granular material behaves as a dry one and in the quasistatic regime in particular (Fig. 6a).

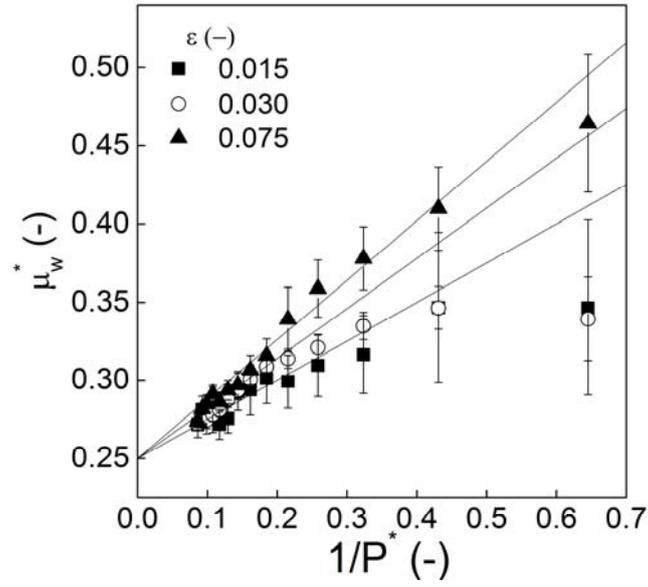

Figure 8: Linear increase of $\mu^*_{w,0}$ with $1/P^*$ for different liquid content.

Figure 9 shows that the macroscopic cohesion, which is the slope of the linear increase of $\mu^*_{w,0}$ vs. $1/P^*$, is a growing function of the liquid content. Furthermore, a fair agreement between experiments and numerical simulations is found. More complete experimental results are presented in another paper (Badetti et al. 2018), in which numerical approaches in the quasistatic limit, are also considered.

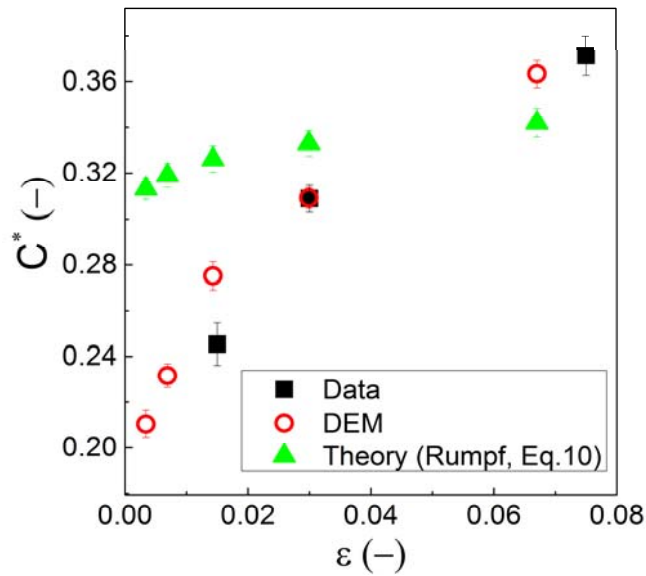

Figure 9: Dimensionless cohesion versus liquid content, as measured in experiments and DEM simulations, and as predited by the Rumpf expression in Eq. 10.

This result is consistent with the cohesion in static shear strength experiments conducted on glass beads by Richefeu et al. (Richefeu et al. 2006) and the relation between the cohesion and the liquid content seems to follow a similar trend as that for the elastic modulus and the liquid content (Møller and Bonn 2007; Fall et al. 2014). In the *pendular* regime, cohesive forces increase with the increasing liquid bridge volume (Chateau et al. 2002).

We now discuss a possible estimation way to predict the *friction law* of wet granular material applying the Rumpf's model (Pietsch et al. 1969). According to Rumpf's theory, values of cohesion $C^*$ of wet granular materials in the *pendular* regime could be estimated with the following expression (Gröger et al. 2003; Richefeu et al. 2006; Khamseh et al. 2015):

$$C^* = \mu_d^* \frac{Z\phi_S}{\pi} \qquad (Eq.10)$$

where $Z$ is the wet coordination number.

Eq. 10 correctly predicts the macroscopic dimensionless cohesion for the higher liquid contents in the investigated range (Fig. 9), but overestimates it at lower liquid contents, for which it fails to capture the decreasing trend. Note that some constant value of the coordination number has to be chosen in (Eq. 10) and the choice made for the data points shown in Figure 9 is explained and discussed in (Badetti et al. 2018), where a more sophisticated estimation of $C^*$, that comes closer to the experimental and numerical values for small liquid content, is also described.

Furthermore, taking thus the coordination number roughly constant of $I$ and $P^*$ (Khamseh et al. 2015), Eq. 1 implies that the theoretical friction coefficient $\mu_{w\_th}^*$ can be written as:

$$\mu_{w\_th}^*\left(I, P^*\right) = \mu_d^*\left(I\right)\left[1 + \frac{Z\varphi_S\left(I, P^*\right)}{\pi P^*}\right] \qquad (Eq.11)$$

From the experimental measurements of the friction coefficient of dry grains and of the solid fraction of the wet sample for a given reduced pressure, Figure 9 shows that (Eq. 11) provides a good estimation of the friction coefficient of the unsaturated wet granular material for different liquid content and for a coordination number set to $Z = 6.2$.

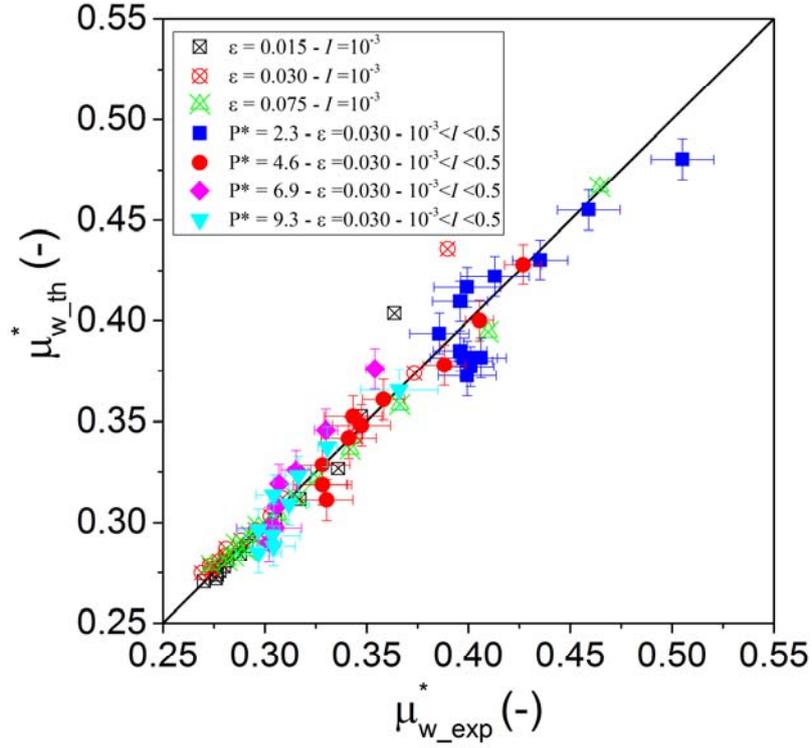

Fig. 10: Predicted values of the friction coefficient using Eq. 11 versus the measured ones, in the quasistatic and inertial flow regimes and for all liquid contents.

IV. CONCLUSION

This paper examines experimentally the transition from the yielding to the flowing behavior of unsaturated wet granular materials. Attention was focused on homogeneously sheared assemblies of particles in an annular shear cell. From well controlled rheometric measurements (under imposed confining normal stress and applied shear rate) and using two simple characterization tools, the internal state of wet granular materials is described. Summarizing, we find that there is a pronounced effect of the addition of small amounts of wetting liquid to dry granular matter. It is observed that the conventional $\mu(I)-$ rheology must be modified to take into account other factors such as cohesion as already observed from DEM simulations (Berger et al. 2015; Khamseh et al. 2015). Compared to dry granular materials, the wet ones exhibit a similar behavior, however, which is dependent on the reduced pressure $P^*$. While the macroscopic friction coefficient increases when $P^*$ decreases, the solid volume fraction is a decreasing function of $P^*$. Such a strong influence of cohesive forces contrasts with the 2D numerical results of (Rognon et al. 2008; Berger et al. 2015), in which similar deviations between cohesionless and cohesive systems are not observed until $P^*$ decreases to much lower values, of order 0.3.

Additionally in the quasistatic regime, a simple effective stress approach may quantitatively account for the shear resistance trend in good approximation as long as $P^* \geq 2$. The Mohr-Coulomb criterion approximately describes critical states in the same reduced pressure range, but is no longer applicable at lower $P^*$. Also results of DEM simulations are shown to agree quantitatively with experimental ones, provided the right value $\mu \approx 0.09$ is given to the intergranular friction coefficient, as identified from the macroscopic properties of the dry material. Note that one limitation of the DEM model is its inability to deal with liquid contents exceeding the pendular regime. Numerical models for higher liquid contents, resorting, e.g., to a lattice Boltzmann discretization of the interstitial liquid, are currently being developed (Roux 2015; Richefeu et al. 2016).

Further use of X-ray tomography enabled investigations of the microstructure. It is found that, for the explored range of liquid content, samples stay homogenous with however the presence of a multitude complex morphologies far from simple capillary bridges. We also observed that shearing tends to reduce the number of these large liquid morphologies to the cost of simple liquid bridges. This important result seems to explain the concordance between experimental and numerical measurements. Indeed, the numerical model is restricted to the *pendular state*, in which the liquid phase is completely discontinuous and no coalescence occurs between liquid bridges.

We believe that the experimental techniques described in this work can be successfully used as a means of investigating the properties of wet cohesive granular materials flows. But as we consider dynamics, there must also be viscous interactions between moving grains, and even the liquid inertia must become important for fast dynamics (Mani et al. 2013). Indeed, the liquid content and the hysteresis for liquid bridge creation and rupture between approaching or receding grain pairs have been observed to significantly affect the rheology in simulations (Roux 2015). Unfortunately, such dynamical effects have been difficult to characterize with conventional rheometric techniques. However, in an ongoing project, we plan to image the evolution of the microstructure under shear with a shear device inserted into MRI imager, enabling local measurements of granular solid fraction and liquid content. A direct access to the velocity or strain field was shown to be crucial to a correct measurement of rheological laws for materials in which strains tend to localize in shear zones (Fall et al. 2015a).

**ACKNOWLEDGMENTS**

We thank D. Hautemayou and C. Mézière for technical helps for the measurements. We gratefully acknowledge financial support from the Agence Nationale de la Recherche (Grant No. ANR-16-CE08-0005-01). The Laboratoire Navier microtomograph has been acquired with the financial support of Region Ile-de-France (SESAME program) and F2M (Fédération Française de Mécanique).

**REFERENCES**

Andreotti B, Forterre Y, Pouliquen O (2013) Granular Media; Between Fluid and Solid. Cambridge Univ Press 55:ISBN-13: 978-1107034792. doi: 10.1080/00107514.2014.885579

Badetti M, Fall A, Chevoir F, Roux J-N (2018) Shear strength of wet granular materials: macroscopic cohesion and effective stress. https://arxiv.org/abs/180208172.

Berger N, Azéma E, Douce J-F, Radjai F (2015) Scaling behaviour of cohesive granular flows. Europhys Lett 112:64004. doi: 10.1209/0295-5075/112/64004

Bocquet L, Charlaix E, Ciliberto S, Crassous J (1998) Moisture-induced ageing in granular media and the kinetics of capillary condensation. Nature 396:735–737. doi: 10.1038/25492

Bocquet L, Charlaix É, Restagno F (2002) Physics of humid granular media. Comptes Rendus Phys 3:207–215. doi: 10.1016/S1631-0705(02)01312-9

Boyer F, Guazzelli É, Pouliquen O (2011) Unifying suspension and granular rheology. Phys Rev Lett 107:1–5. doi: 10.1103/PhysRevLett.107.188301

Bruchon JF, Pereira JM, Vandamme M, et al (2013) Full 3D investigation and characterisation of capillary collapse of a loose unsaturated sand using X-ray CT. Granul Matter 15:783–800. doi: 10.1007/s10035-013-0452-6

Chateau X, Moucheront P, Pitois O (2002) Micromechanics of Unsaturated Granular Media. J Eng Mech 128:856–863. doi: 10.1061/(ASCE)0733-9399(2002)128:8(856)

Chou SH, Hsiau SS (2010) Experimental analysis of the dynamic properties of wet granular matter. Powder Technol 197:222–229. doi: 10.1016/j.powtec.2011.09.010

Cleaver JAS, Nedderman RM, Thorpe RB (2000) Accounting for granular material dilation during the operation of an annular shear cell. Adv Powder Technol 11:385–399. doi:

https://doi.org/10.1163/156855200750172015

Coste C (2004) Shearing of a confined granular layer: Tangential stress and dilatancy. Phys Rev E 70:1–11. doi: 10.1103/PhysRevE.70.051302

Coster M, Chermant J-L (2001) Image analysis and mathematical morphology for civil engineering materials. Cem Concr Compos 23:133–151. doi: 10.1016/S0958-9465(00)00058-5

Da Cruz F, Emam S, Prochnow M, et al (2005) Rheophysics of dense granular materials: Discrete simulation of plane shear flows. Phys Rev E 72:21309. doi: 10.1103/PhysRevE.72.021309

Day PR, Holmgren GG (1952) Microscopic Changes in Soil Structure During Compression1. Soil Sci Soc Am J 16:73–77. doi: 10.2136/sssaj1952.03615995001600010022x

Estrada N, Lizcano A, Taboada A (2010) Simulation of cemented granular materials. I. Macroscopic stress-strain response and strain localization. Phys Rev E. doi: 10.1103/PhysRevE.82.011303

Fall A, Bertrand F, Hautemayou D, et al (2015a) Macroscopic discontinuous shear thickening versus local shear jamming in cornstarch. Phys Rev Lett 114:1–5. doi: 10.1103/PhysRevLett.114.098301

Fall A, Ovarlez G, Hautemayou D, et al (2015b) Dry granular flows : Rheological measurements of the μ ( I ) -rheology. J Rheol (N Y N Y) 59:1065–1080. doi: 10.1122/1.4922653

Fall A, Paredes J, Bonn D (2010) Yielding and shear banding in soft glassy materials. Phys Rev Lett. doi: 10.1103/PhysRevLett.105.225502

Fall A, Weber B, Pakpour M, et al (2014) Sliding friction on wet and dry sand. Phys Rev Lett. doi: 10.1103/PhysRevLett.112.175502

Fiscina JEE, Pakpour M, Fall A, et al (2012) Dissipation in quasistatically sheared wet and dry sand under confinement. Phys Rev E 86:20103. doi: 10.1103/PhysRevE.86.020103

GDR Midi (2004) On dense granular flows. Eur Phys J E Soft Matter 14:341–65. doi: 10.1140/epje/i2003-10153-0

Geromichalos D, Kohonen MM, Mugele F, Herminghaus S (2003) Mixing and Condensation in a Wet Granular Medium. Phys Rev Lett 90:168702. doi: 10.1103/PhysRevLett.90.168702

Ghezzehei TA, Or D (2000) Dynamics of soil aggregate coalescence and theological processes. Water Resour Res 36:367–379.

Gilabert FA, Roux JN, Castellanos A (2008) Computer simulation of model cohesive powders: Plastic

consolidation, structural changes, and elasticity under isotropic loads. Phys Rev E 78:31305. doi: 10.1103/PhysRevE.78.031305

Gray J. MNT, Edwards AN (2014) A depth-averaged µ(I)-rheology for shallow granular free-surface flows. J Fluid Mech 755:503–534. doi: 10.1017/jfm.2014.450

Gröger T, Tüzün U, Heyes DM (2003) Modelling and measuring of cohesion in wet granular materials. Powder Technol 133:203–215. doi: 10.1016/S0032-5910(03)00093-7

Hatano T (2007) Power-law friction in closely packed granular materials. Phys Rev E 75:1–4. doi: 10.1103/PhysRevE.75.060301

Herminghaus S (2005) Dynamics of wet granular matter. Adv Phys 54:221–261. doi: 10.1080/00018730500167855

Iveson SM, Litster JD, Hapgood K, Ennis BJ (2001) Nucleation, growth and breakage phenomena in agitated wet granulation processes: A review. Powder Technol 117:3–39. doi: 10.1016/S0032-5910(01)00313-8

Jop P, Forterre Y, Pouliquen O (2006) A constitutive law for dense granular flows. Nature 441:727–730. doi: 10.1038/nature04801

Khamseh S, Roux JN, Chevoir F (2015) Flow of wet granular materials: A numerical study. Phys Rev E 92:22201. doi: 10.1103/PhysRevE.92.022201

Kuwano O, Ando R, Hatano T (2013) Crossover from negative to positive shear rate dependence in granular friction. Geophys Res Lett 40:1295–1299. doi: 10.1002/grl.50311

Lefebvre G, Jop P (2013) Erosion dynamics of a wet granular medium. Phys Rev E. doi: 10.1103/PhysRevE.88.032205

Lemaître A, Roux JN, Chevoir F (2009) What do dry granular flows tell us about dense non-Brownian suspension rheology? In: Rheologica Acta. pp 925–942

Mani R, Kadau D, Herrmann HJ (2013) Liquid migration in sheared unsaturated granular media. Granul Matter 15:447–454. doi: 10.1007/s10035-012-0387-3

Maugis D (1987) Adhesion of elastomers: Fracture mechanics aspects. J Adhes Sci Tec 1:105. doi: 10.1163/156856187X00120

Mikami T, Kamiya H, Horio M (1998) Numerical simulation of cohesive powder behavior in a fluidized

bed. Chem Eng Sci 53:1927–1940. doi: 10.1016/S0009-2509(97)00325-4

Mitarai N (2012) Granular flow: dry and wet. Eur Phys J Spec Top 204:5–17.

Mitarai N, Nori F (2006) Wet granular materials. Adv Phys 55:1–45. doi: DOI 10.1080/00018730600626065

Møller PCF, Bonn D (2007) The shear modulus of wet granular matter. Europhys Lett 80:38002. doi: 10.1209/0295-5075/80/38002

Ovarlez G, Mahaut F, Deboeuf S, et al (2015) Flows of suspensions of particles in yield stress fluids. J Rheol (N Y N Y) 59:1449–1486. doi: 10.1122/1.4934363

Pakpour M, Habibi M, Møller P, Bonn D (2012) How to construct the perfect sandcastle. Sci Rep. doi: 10.1038/srep00549

Paredes J, Shahidzadeh-Bonn N, Bonn D (2011) Shear banding in thixotropic and normal emulsions. J Phys Condens Matter 23:284116. doi: 10.1088/0953-8984/23/28/284116

Pierrat P, Agrawal DK, Caram HS (1998) Effect of moisture on the yield locus of granular materials: Theory of shift. Powder Technol 99:220–227. doi: 10.1016/S0032-5910(98)00111-9

Pietsch W, Hoffman E, Rumpf H (1969) Tensile strength of moist agglomerates. Ind Eng Chem Prod Res Dev 8:58–62. doi: 10.1021/i360029a009

Richefeu V, El Youssoufi MS, Radjaï F (2006) Shear strength properties of wet granular materials. Phys Rev E 73:1–11. doi: 10.1103/PhysRevE.73.051304

Richefeu V, Radjai F, Delenne JY (2016) Lattice Boltzmann modelling of liquid distribution in unsaturated granular media. Comput Geotech 80:353–359. doi: 10.1016/j.compgeo.2016.02.017

Rognon PG, Roux J-N, Naaïm M, Chevoir F (2008) Dense flows of cohesive granular materials ¨. J Fluid Mech 596:21–47. doi: 10.1017/S0022112007009329

Rognon PG, Roux J-N, Wolf D, et al (2006) Rheophysics of cohesive granular materials. Europhys Lett 74:644. doi: 10.1209/epl/i2005-10578-y

Roux J-N (2015) A numerical toolkit to understand the mechanics of partially saturated granular materials. J Fluid Mech 770:1–4. doi: 10.1017/jfm.2015.66

S. B. Savage and M. Sayed (1984) Stresses developed by dry cohesionless granular materials sheared

in an annular shear cell. J Fluid Mech 53142:391–430. doi: 10.1017/CBO9781107415324.004

Saingier G, Sauret A, Jop P (2017) Accretion Dynamics on Wet Granular Materials. Phys Rev Lett. doi: 10.1103/PhysRevLett.118.208001

Samadani A, Kudrolli A (2000) Segregation transitions in wet granular matter. 2–5. doi: 10.1103/PhysRevLett.85.5102

Samadani A, Kudrolli A (2001) Angle of repose and segregation in cohesive granular matter. Phys Rev E 64:51301. doi: 10.1103/PhysRevE.64.051301

Scheel M, Geromichalos D, Herminghaus S (2004) Wet granular matter under vertical agitation. J Phys Condens Matter 16:S4213–S4218. doi: 10.1088/0953-8984/16/38/033

Scheel M, Seemann R, Brinkmann M, et al (2008) Morphological clues to wet granular pile stability. Nat Mater 7:189–193. doi: 10.1038/nmat2117

Scholtès L, Chareyre B, Nicot F, Darve F (2009) Micromechanics of granular materials with capillary effects (DOI:10.1016/j.ijengsci.2008.07.002). Int J Eng Sci 47:1460–1471. doi: 10.1016/j.ijengsci.2009.10.003

Schwarze R, Gladkyy A, Uhlig F, Luding S (2013) Rheology of weakly wetted granular materials: A comparison of experimental and numerical data. Granul Matter 15:455–465. doi: 10.1007/s10035-013-0430-z

Semprebon C, Scheel M, Herminghaus S, et al (2016) Liquid morphologies and capillary forces between three spherical beads. Phys Rev E 94:12907. doi: 10.1103/PhysRevE.94.012907

Soulié F, El Youssoufi MS, Cherblanc F, Saix C (2006) Capillary cohesion and mechanical strength of polydisperse granular materials. Eur Phys J E 21:349–357. doi: 10.1140/epje/i2006-10076-2

Staron L, Lagrée PY, Josserand C, Lhuillier D (2010) Flow and jamming of a two-dimensional granular bed: Toward a nonlocal rheology? Phys Fluids 22:1–10. doi: 10.1063/1.3499353

Tegzes P, Vicsek T, Schiffer P (2003) Development of correlations in the dynamics of wet granular avalanches. Phys Rev E 67:51303. doi: 10.1103/PhysRevE.67.051303

Than V Du, Khamseh S, Tang AM, et al (2017) Basic Mechanical Properties of Wet Granular Materials : A DEM Study. J Eng Mech 143:C4016001. doi: 10.1061/(ASCE)EM.1943-7889.0001043.